\documentclass[conference]{IEEEtran}
\usepackage{epsf}
\usepackage{times}
\usepackage{epsfig}
\usepackage{epstopdf}
\usepackage{graphicx}
\usepackage{flushend}
\usepackage{amsxtra}
\usepackage{multirow}
\usepackage{mathtools}
\usepackage{subcaption}
\usepackage{multirow}
\usepackage{comment}
\usepackage[normalem]{ulem}
\usepackage{color}
\usepackage{amsmath}
\usepackage{amssymb}

\usepackage[normalem]{ulem}
\usepackage{tabu}
\usepackage[dvisnames]{xcolor}
\usepackage{multicol}
\usepackage{cite}
\usepackage{caption}
\begin{document}
\title{Optimal Team Recruitment Strategies for Collaborative Mobile Crowdsourcing Systems}

\author{\IEEEauthorblockN{Aymen Hamrouni$^1$, Hakim Ghazzai$^1$, Turki Alelyani$^2$, and Yehia Massoud$^1$}
\IEEEauthorblockA{\small $^1$School of Systems  \& Enterprises, Stevens Institute of Technology, Hoboken, NJ, USA\\
Email: \{ahamroun, hghazzai, ymassoud\}@stevens.edu} 
$^2$College of Computer Science and Information Systems, Najran University, Najran, Saudi Arabia \\
Email: tnalelyani@nu.edu.sa}
\maketitle

\thispagestyle{empty}

\begin{abstract}
\boldmath{The wide spread of mobile devices has enabled a new paradigm of innovation called Mobile Crowdsourcing (MCS) where the concept is to allow entities, e.g., individuals or local authorities, to hire workers to help from the crowd of connected people, to execute a task or service. Some complex tasks require the collaboration of multiple workers to ensure its successful completion. In this context, the task requester needs to hire a group of socially connected and collaborative workers that, at the same time, have sufficient skills to accomplish the task. In this paper, we develop two recruitment strategies for collaborative MCS frameworks in which, virtual teams are formed according to four different criteria: level of expertise, social relationship strength, recruitment cost, and recruiter's confidence level. The first proposed strategy is a platform-based approach which exploits the platform knowledge to form the team. The second one is a leader-based approach that uses team members' knowledge about their social network (SN) neighbors to designate a group leader that recruits its suitable team. Both approaches are modeled as integer linear programs resulting in optimal team formation. Experimental results show a performance trade-off between the two virtual team grouping strategies when varying the members SN edge degree. Compared to the leader-based strategy, the platform-based strategy recruits a more skilled team but with lower SN relationships and higher cost.}
\end{abstract}


\section{Introduction}
\label{Sec1a}

Mobile Crowdsourcing (MCS) is the act of outsourcing sensing tasks traditionally performed by employees or contractors to an undefined large group of dynamic Internet population or cyber community through an open or targeted call. It harnesses the power of built-in sensors in mobile devices (i.e., smartphones, tablets, smart devices) and allows Internet-of-things (IoT) devices to establish relationships and cooperate together to complete specific sensing and data collection tasks without requiring pre-deployed dedicated infrastructure~\cite{7491206,7538981}.

In a typical MCS architecture~\cite{article5}, there are three main agents: task requesters, task workers, and the cloud platform hosting the main framework. The task requester, which may be a human carrying a smartphone as well as a machine (e.g., autonomous vehicle), provides its task information and requirements to the cloud platform to execute a certain task, e.g., collecting photos or sensing data. The platform then uses these criteria to recruit suitable workers and provides them with task requirements. Most of the existing MCS approaches tackle simple tasks such as photo collection ~\cite{8982179} or improving the labeling accuracy~\cite{7875084}. In each of these MCS tasks, selected workers are asked to achieve what is necessary independently of each other and the final result is combined from their partial results to produce the overall outcome. To this end, most of the recent MCS researchers, for example, Cheng et al.~\cite{7446292} focused on optimizing the recruitment process by hiring skilled workers for each task such that they can fulfill the tasks' requirements and provide suitable results.

However, in many other MCS applications, the set of tasks, also called projects, are very complex and the success of their completion depends not only on the expertise of their selected workers but also on how efficiently these workers can work together as a team~\cite{8488386}. This could be, for example, the case of a MCS framework that helps build a virtual search party of smartphone users to find lost items, pets, or persons, as well as returning them. The enlisted workers are divided into convenient groups, using specific criteria required by the requester, and are asked to collaborate together for the search by providing up-to-the minute reports about any updates. If the collaboration somehow fails for one reason or another, the job cannot be achieved successfully. Therefore, besides providing the required skills, the successful completion of the project is very sensitive to the way team members collaborate and communicate. This type of MCS is called Collaborative MCS (CMCS).

CMCS is a teamwork-based paradigm where the set of workers, often with diverse and complementary skills, form groups and work together to complete complex tasks~\cite{2222,Mass1912,7806269}.  In traditional crowdsourcing applications, workers are recruited and asked to complete the same task independently of each other and without any contact. Recent studies have begun to address the need to consider recruiting a team of workers in MCS ~\cite{article,7248382,AAAI1612106}. In fact, some approaches, such as~\cite{116}, focused on dividing complex tasks into flows of simple sub-tasks and allocating these sub-tasks to a team of workers. At the end, the partial results are combined to produce the overall outcome. These approaches have focused only on the expertise of recruited team members and did not consider the interaction within members. Other approaches limited their focus on team formation in Social Networks (SNs) and proposed a solution to hire teams with good social relationships indifferent of the members' levels of expertise~\cite{8488386,article7,article8}. 
 
In this paper, we complement these studies and investigate two CMCS recruitment techniques that combine social interaction with expertise and consider four key recruitment metrics: required skills, social relationships, budget allocation, and recruiter confidence level. The first recruitment approach is a platform-based strategy in which the CMCS platform itself is responsible for forming the entire team based on its knowledge about the workers SNs and its attributes (e.g., profile, history, experience, previous performance, reliability). The second one is a leader-based strategy in which the cloud platform selects a worker as a leader to which it delegates the team formation procedure. The chosen leader recruits team members based on its knowledge about other workers in its SN vicinity (e.g., social incentive mechanism, \mbox{man-to-man}, friendship). In layman's terms, we propose:\\
$\bullet$ A platform-based recruitment strategy that exploits the knowledge of the platform towards workers and recruits suitable team members.\\
$\bullet$ A leader-based recruitment strategy that uses the knowledge of an appointed leader by the platform to recruit the rest of the team. \\
The common goal of both of these recruitment strategies is to form not only, skilled teams, but also socially connected. The platform-based approach utilizes the overall knowledge of the platform about the workers' attributes and SNs. However, although the platform typically has a global view of all available workers attributes, its knowledge is limited with a low accuracy and precision. To this end, we introduce the leader-based approach which makes use of the workers knowledge about other workers' skills and profiles in their vicinity. This approach relies on locality and includes the leader's knowledge who is usually better informed about the workers in its SN vicinity than the platform. Both CMCS recruitment strategies are modeled as Integer Linear Programs (ILP). Simulation results show a performance trade-off between the two recruitment strategies when varying the workers SN edge degree. Compared to the leader-based strategy, the platform-based strategy recruits a more skilled team but with lower SN relationships and higher cost.

\section{CMCS Model}\label{Sec2}
\begin{figure}[t]
    \centering
    \vspace{0.3cm}
    \includegraphics[width=9cm]{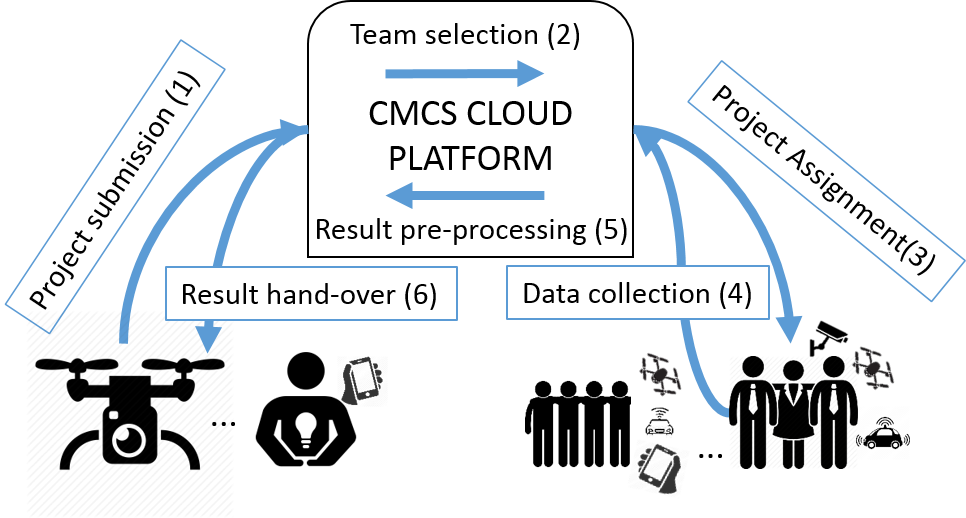}
    \caption{Work-flow of the typical CMCS platform.}
    \label{cmcsplatform}
\end{figure}

In a typical CMCS system, there are two external parties which interact with the platform. As illustrated in Fig.~\ref{cmcsplatform}, these actors are the project initiator (e.g., local authority, weather company, mobile user, etc.) and the available workers (e.g., humans with smartphones, autonomous vehicles, sensors, etc). When it needs services, the project initiator submits its CMCS project $p$, having $\mathcal S_p$ as a set of required skills (e.g., expertise for humans, device specifications, etc.), to the platform. The latter is usually  a centralized computing architecture responsible of recruiting a suitable team.

We denote by $\mathcal J$ the set of $J$ workers registered in the CMCS platform where $\mathcal J=\{1, \dots, J\}$. Given the set $\mathcal S=\{1,\dots, S\}$ of all $S$ possible skills in the system, we define the logical skill quantity for a project $p$ by $Q_p(k),\,k\ \in \mathcal S$ where $Q_p(k)=1$ if the skill $k$ is required by project $p$ and $Q_p(k)=0$ otherwise. Hence, the skills set required by the project $\mathcal S_p=\{k \in \mathcal S/Q_p(k)=1\}$. Each worker $j \in \mathcal J$ has a degree of expertise in skill $k \in \mathcal S$ denoted by $S_{jk}$ where $0\leq S_{jk} \leq 1$. The term $S_{jk}$ represents the actual expertise value of skill $k$ that worker $j$ have and it is interpreted as follows: $S_{jk} \leftarrow 1$ means that the worker $j$ is an expert in skill $k$. Otherwise, $S_{jk}\rightarrow 0$.  We assume that a recruiter $i$, which can be a worker as well as the platform itself, does not perfectly know the degree of the skill $k$ of each worker. Instead, it knows an estimated value expressed as follows: $\hat{S}^i_{jk}=S_{jk}+\tilde{S}^i_{j}$, where $\tilde{S}^i_{j}$ is a skill error made by the recruiter $i$ given its knowledge about worker $j$. Let $\hat{S}_j=\{\hat{S}_{j1}, \dots, \hat{S}_{jM}\}$ be the set of skills  provided by worker $j$. We suppose that each recruited worker can only contribute with one required skill. Consequently, for a project $p$ having as a skill set $\mathcal S_p$, the number of team members must be $|\mathcal S_p|$.  

To execute a task with skill $k$, a worker $j$ may request a certain cost denoted by ~$C_{jk}$. 
We suppose that the workers in the platform form a SN modeled as an undirected weighted graph $\mathcal G(\mathcal J,\mathcal E)$. Every vertex of $\mathcal G$ corresponds to a worker $j \in \mathcal J$ while the set of edges $ \mathcal E$ represents the SN relationships between the workers. Initially, we only consider the edges connecting a pairwise of workers that can directly communicate and collaborate and we associate the value $1$ to their weights. Then, the edges between the remaining pairwise of vertices, e.g., $(j,j')$, which are not directly connected are given a weight computed using the shortest number of hops, denoted by $n^{hops}_{jj'}$, needed for one of the pairwise vertices to reach the other. Hence, the graph $\mathcal G$ is converted into a complete graph where all vertices are directly connected and the values of the edges' weights indicate the social relationships levels between each pair of workers. The real values on each edge between two workers $j$ and $j'$ are given as: $R_{jj'}=\frac{1}{1+n^{hops}_{jj'}}$. We also assume that the recruiter $i$ does not perfectly knows the relationships degree between workers in the SN. Instead, it knows an estimated value expressed as:
$\hat{R}_{jj'}=R_{jj'}+\frac{(\tilde{R}^i_{j}+\tilde{R}^{i}_{j'})}{2}$, where $\tilde{R}^i_{j}$ and $\tilde{R}^i_{j'}$ represent relationships error made by recruiter $i$ given its knowledge about workers $j$ and $j'$, respectively. Both the noises $\tilde{R}^i_{j}$ and $\tilde{S}^i_{j}$ are modeled as a probability distribution with a variance $U^i_j$ that reflects the confidence level of the recruiter $i$ towards worker $j$. If an isolated sub-graph exists, then the weights connecting a node of this sub-graph to other external vertices is a pure noise (i.e., $n_{hops}\rightarrow \infty$).

The efficiency of worker $j$ chosen by recruiter $i$ to contribute with skill $k$ is written as follows: 
\begin{align}
E^i_{j,k}= \eta_1\frac{ \hat{S}^i_{j,k}}{\bar{S}}  - \eta_2 \frac{ U^{i}_j}{\bar{U}}  
-   \eta_3 \frac{C_{j,k}}{\bar{ C }}  + \frac{\eta_4}{|\mathcal T|-1} \sum_{j'\in \mathcal T \backslash\ \{j\}} \frac{\hat{R}_{jj'}}{\bar{R}},
\label{ppp}
\end{align}
where $\mathcal T$ is the set of hired workers in the formed team. The efficiency expression which the platform aims to maximize is established using the four key metrics. The quantities $\bar{X}$ in \eqref{ppp} are introduced for normalization purposes so that the four key metrics have the same order of magnitude in the efficiency expression. The weights $\eta_t$, with $t \in \{1,2,3,4\}$ and $\sum_{t=1}^{4} \eta_t=1$, indicate the recruiter's recruitment strategy. For example, situations where the project requester only cares about its task being completed by the workers having the highest skills (i.e., $\eta_1=1$, and $\eta_2, \eta_3, \eta_4$ are set to 0). If the recruiter is looking for a reduced cost-effective team, a higher value of $\eta_3$ is recommended.

\section{Problem Formulation}
\label{sec3}
In this section, we formulate a general optimization problem adequate for both of the recruitment strategies. Initially, we start by defining the required decision variables. Then, we introduce the optimization problems while specifying the common constraints for the two approaches and also the ones that are specific to each recruitment strategy.

Let $\mathcal I$ be the set of possible team recruiters, which can be defined as follows:
\begin{align} \label{gamma}
\mathcal I= 
     \begin{cases}
       \text{\{0\},} &\,\text{if the recruiter is the platform,}\\
      \mathcal J, &\,\text{if the recruiter is a worker.}
     \end{cases}
     \end{align}
In order to assign to the recruiter the chosen workers for a project $p$ and their contributed skill $s_k$, we introduce a binary decision variable $x^i_{jk}$ defined as follows:
\begin{align} \label{x}
x^i_{jk}= &
     \begin{cases}
       \text{1,} &\,\text{if recruiter $i$ selects worker $j$ to contribute}\\
       &\text{in project $p$ with skill $k$,}\\
       \text{0,} &\,\text{otherwise,} \\ 
     \end{cases}\notag\\
 &\hspace{2cm}\forall \, (i,j) \in \mathcal I \times \mathcal J, \forall \, k \in \mathcal S_p.
     \end{align}
Depending on the recruitment strategy, the index $i$ can take either the value of $\{0\}$ for the platform-based approach or any value in $\mathcal J$ for the leader-based approach. Furthermore, this index indicates the identity of the recruiter $i$ to specify to the optimizer which entity's knowledge about the workers attributes it will base its recruitment decision.

Another binary decision variable $v_{jj'}$ is introduced to consider the social relationships between the project team mates. This variable is presented as follows:
\begin{align} \label{v}
& v_{jj'}= 
     \begin{cases}
       \text{1,} &\,\text{if workers $j$ and $j'$ are hired within the project,}\\
       \text{0,} &\,\text{otherwise,} \\ 
     \end{cases}\notag\\
 &\hspace{2cm}\forall \, (j,j') \in \mathcal J \times \mathcal J.
     \end{align}
The variable $v_{jj'}$ indicates that all the positive 2-tuple combinations of the chosen team members. Its value can be computed using the following expression:
\begin{align}
&v_{jj'}= \sum_{i \in \mathcal I }\sum_{k \in \mathcal S_p}^{} x^i_{jk} \land \sum_{i \in \mathcal I } \sum_{k \in \mathcal S_p}^{} x^i_{j'k}, \forall \, (j,j') \in \mathcal J \times \mathcal J, \label{valueofv}
\end{align}
where the symbol ($\land$) represents the logical operator \textit{AND}.

The objective of this paper is to hire the most suitable team to complete a CMCS project $p$. To this end, we introduce a general team formation optimization problem for both approaches and we define it as follows:
\begin{align}
\text{(P):} & \underset{ \underset{ v_{jj'} \in \{0,1\}}{x^i_{jk} \in \{0,1\}}}{\text{ maximize }}
\sum_{i \in \mathcal I } \sum_{j \in \mathcal J}^{}  \sum_{k \in \mathcal S_p}^{} x^i_{jk} \bigg [ \eta_1 \frac{ \hat{S}^i_{j,k}}{\bar{S}}  - \eta_2 \frac{  U^{i}_j }{\bar{U}}
-  \eta_3 \frac{  C_{j,k}}{\bar{C}}  \bigg  ] \notag \\ 
&\hspace{2cm} + \frac{\eta_4}{|\mathcal S_p|-1} \sum_{j \in \mathcal J}^{}  \sum_{j'  \in \mathcal J}^{} v_{jj'}  \frac{R_{jj'}}{\bar{R}},  \notag \\ 
&\text{subject to:} \notag \\ 
&\hspace{-0.25cm}\sum_{i \in \mathcal I } \sum_{k \in \mathcal S_p}^{}x^i_{jk}\leq 1, \, \forall \, j \in \mathcal J,\label{1}
\end{align}
\begin{subequations}
\begin{align}
&\sum_{ i \in \mathcal I } \sum_{k \in \mathcal S_p}^{} ( x^i_{jk}+x^i_{j'k})  \geq  2 \times v_{jj'}, \, \forall \, (j,j') \in \mathcal J \times  \mathcal J, \label{3}\\
&v_{jj'}  \geq \sum_{ i \in \mathcal I } \sum_{k \in \mathcal S_p}^{} ( x^i_{jk}+x^i_{j'k}) -1, \, \forall \, (j,j') \in \mathcal J \times \mathcal J, \label{4}
\end{align}
\end{subequations}
\begin{align}
&v_{jj} = 0, \forall \, j \in \mathcal J \text{ and } v_{j'j} = v_{jj'}, \,   \forall \, (j,j') \in \mathcal J \times \mathcal J.\label{6}
\end{align}

The previous constraints are common for both recruitment strategies. The following constraint is specifically added to the platform-based approach:
\begin{equation}
\hspace{-0.4cm}\sum_{i \in \mathcal I } \sum_{j  \in \mathcal J}^{}x^i_{jk} = Q(s_k), \, \forall \, k \in \mathcal S_p,\label{2}
\end{equation}
while the constraints \eqref{leader}-\eqref{8} below are considered in the leader-based approach:
\begin{subequations}
\begin{align}
&\sum_{k \in \mathcal S_p}^{} \left( \sum_{j  \in \mathcal J}^{} x^i_{jk} - Q(s_k) \right) \leq M  (1-y^{i}), \forall \,i \in \mathcal J, \label{9}\\
&  \sum_{k \in \mathcal S_p}^{} \left( \sum_{j  \in \mathcal J}^{} x^i_{jk} - Q(s_k) \right) \geq M  (y^{i}-1), \forall \,i\in \mathcal J, \label{10}\\
&\sum_{j  \in \mathcal J}^{} \sum_{k \in \mathcal S_p}^{} x^i_{jk}  \leq M  y^{i}, \forall \,i\in \mathcal J, \label{11}\\
&\sum_{j  \in \mathcal J}^{} \sum_{k \in \mathcal S_p}^{} x^i_{jk}  \geq -M y^{i}, \forall \,i\in \mathcal J, \label{12}
\end{align}
\label{leader}
\end{subequations}
\begin{equation}
\hspace{-2.5cm}\sum_{k \in \mathcal S_p}^{}x^i_{ik} = 1, \forall \, i \in \mathcal J \text{ and } \sum_{i  \in \mathcal J}y^{i} = 1, \, \label{8}
\end{equation}
where $y^i, \forall i \in J$, is a  endogenous binary variable that equals $1$ if worker $i$ is the leader of the hired team and $0$, otherwise. Constraint~\eqref{1} forces each worker to provide at most one skill for the project $p$. The value of $v_{jj'}$ is computed using a product of the decision variables $x^i_{jk}$. Therefore, we use the standard linearization technique and replace $v_{jj'}$ given in~\eqref{valueofv} with the constraints~\eqref{3} and~\eqref{4}.

Constraints~\eqref{6} address the SNs of the workers within the team. In fact, the first term eliminates the case of counting a worker $j$ to be its own co-worker within the hired team. The second term highlights the symmetric relations between vertices in the undirected graph $\mathcal G(\mathcal J, \mathcal E)$ of workers' SNs.

The constraints presented in~\eqref{2} and~\eqref{leader} have analogous goal but they are adapted for each strategy. On one hand, constraint in \eqref{2} is presented for the platform-based approach and ensures that each of the workers within the hired team contributes with a required skill defined by the project. On the other hand, constraints \eqref{9}, \eqref{10}, \eqref{11}, and \eqref{12} are the result of a Big-M method to guarantee that the team leader recruits a team with workers having the required skills. The term $M$ represents the upper bound of $\sum_{j  \in \mathcal J}^{} \sum_{k \in \mathcal S_p}^{} x^i_{jk}$. With the leader-based constraints, we add the first term in constraints \eqref{8} to ensure that the chosen leader is also a team member that contributes in the project with one required skill. To guarantee the uniqueness of the leader, we include the second term in constraints given in \eqref{8}. 

\begin{figure}[t]
    \centering
    \vspace{0.25cm}
    \hspace{-0.35cm}
    \includegraphics[width=8.9cm]{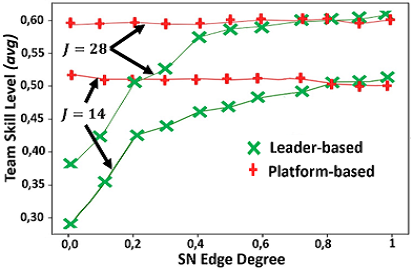}
\caption{Average skills level of the recruited team vs. social network edge degree for the leader-based and the platform-based recruitment strategies with $|\mathcal S_p|=5$ for values of $J$ set to $14$ and $28$.}
\label{fig1}
\end{figure}

This optimization problem in (P) is formulated as an ILP and the solution can be optimally obtained using off-the-shelf software such as Gurobi or CPLEX integrating the branch and bound algorithm and simplex method. Also, note that there are cases where the problem is infeasible, for example when the number of workers $J$ is less than the number of skills $|\mathcal S_p|$. However, in CMCS platforms, this case is unlikely to occur since by definition, in large-scale IoT systems, the value of $J \gg  |\mathcal S_p|$.

\vspace{0.5cm}
\section{Experiments and Evaluation}\label{sec4}

\begin{figure}[t]
    \centering
    \vspace{0.25cm}
   \hspace{-0.35cm}
    \includegraphics[width=8.9cm]{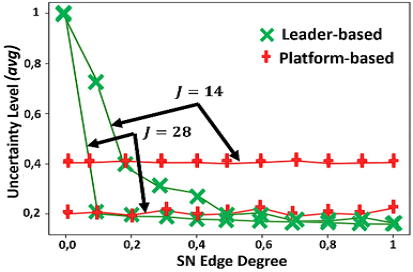}
\caption{Average recruiter's uncertainty level vs. social network edge degree for the leader-based and the platform-based recruitment strategies with $|\mathcal S_p|=5$ for values of $J$ set to $14$ and $28$.   \vspace{0.2cm}}
\label{fig2}

\end{figure}
\begin{figure}[t]
    \centering
    \vspace{0.2cm}
   \hspace{-0.33cm}
    \includegraphics[width=9cm]{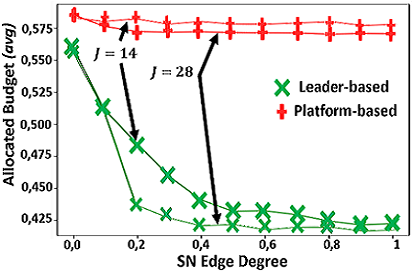}
\caption{Average cost per worker vs. social network edge degree for the leader-based and the platform-based recruitment strategies with $|\mathcal S_p|=5$ for values of $J$ set to $14$ and $28$.  \vspace{0.2cm}}
\label{fig3}
\end{figure}

\begin{figure}[!t]
    \centering
    \hspace{-0.32cm}
    \includegraphics[width=9cm]{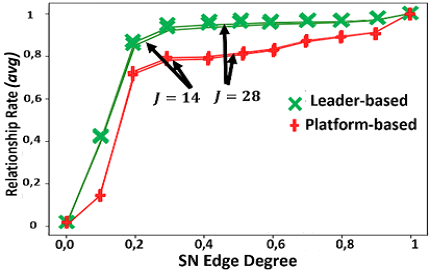}
\caption{Social relationships rate vs. social network edge degree for the leader-based and the platform-based recruitment strategies with $|\mathcal S_p|=5$ for values of $J$ set to $14$ and $28$.  \vspace{0.2cm}}
\label{fig4}
\end{figure}

\begin{figure*}[t]\vspace{0.05cm}
\centering
     \includegraphics[width=15cm]{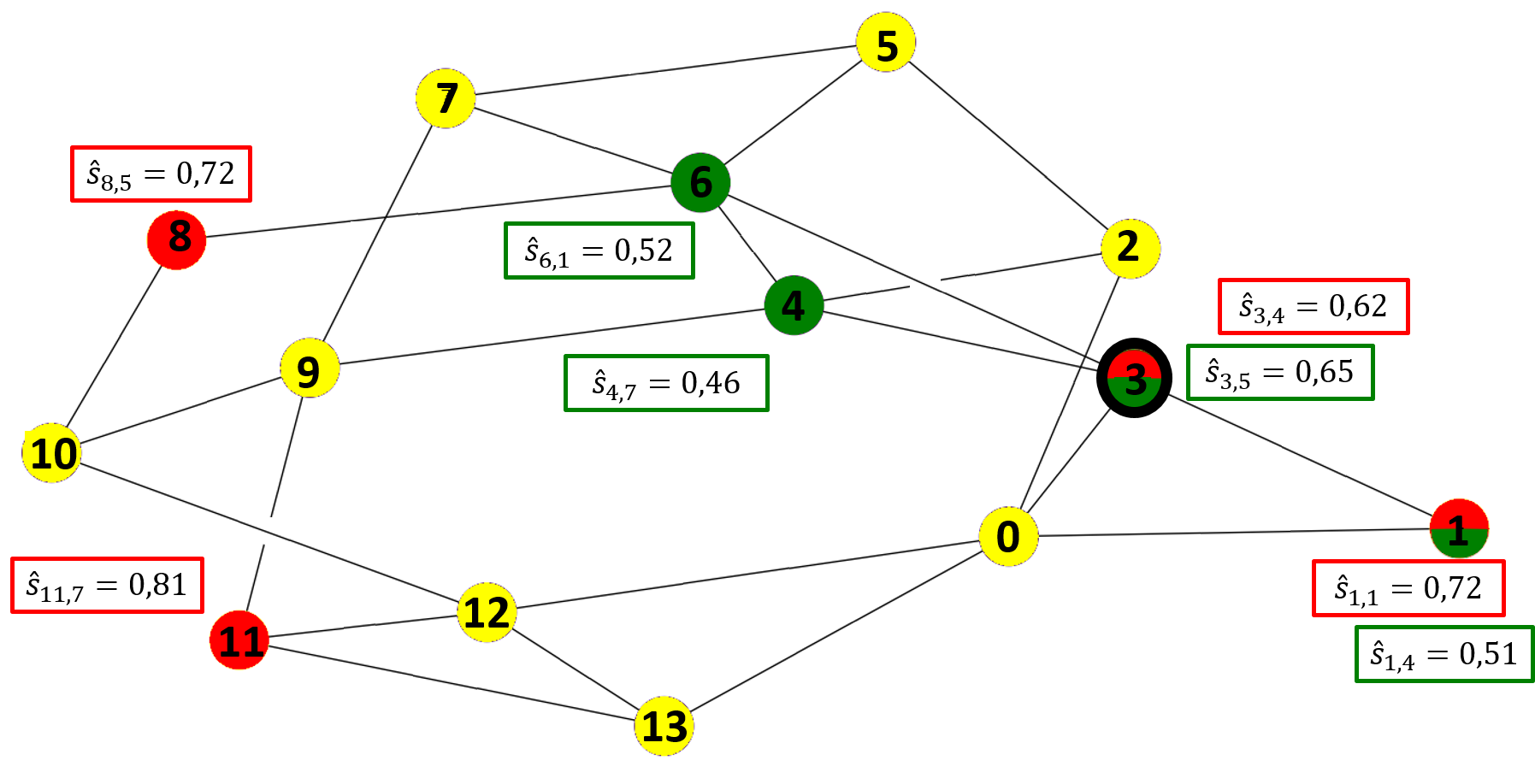}
\caption{An example of the selected teams using both recruitment strategies for an SN graph where $J=14$ and $|\mathcal S_p|=4$. The selected team members are green and red for the leader-based and the platform-based approaches. For the former approach, the team leader is contoured in bold.}
\label{2figures}
\end{figure*}

In order to simulate the team formation process in CMCS, we use a synthetic data with different types of project requirements and workers' skills. We use Watts\text{-}Strogatz model to randomly produce the graph $\mathcal G(\mathcal J,\mathcal E)$ with small-world properties. For the platform-based approach, the values of the error levels on the workers' skills and the workers' relationships are proportionally modeled to the workers history (i.e., workers with more history in the platform has lower uncertainty levels). On the other hand, the uncertainty levels of potential leaders $i$ towards workers $j$ for the leader-based recruitment strategy and the error levels on the workers relationships are proportional and increase with the number of hops between the team leader and other workers. For both approaches, the error is modeled as a zero-mean normal distribution $\sim \mathcal{N}(0,\,0.3^2)$.

In Fig.~\ref{fig1}-\ref{fig4}, we perform a Monte Carlo simulations where $1,000$ realizations of different parameter settings are generated. We evaluate the average four metrics of the selected teams: skills efficiency, recruiter confidence, team cost, and social relationships while varying the edge density of the SN graph $\mathcal G(\mathcal J, \mathcal E)$ for $J=14$ and $J=24$. We adopt a proportional setting and set $\eta_t=0.25$, $\forall t \in \{0,1,2,3\}$. We also set the values of $M$ and $|\mathcal S_p|$ to $8$ and $5$, respectively.

The result of this simulation for $J=14$ shows that, for the leader-based approach, the performances for all metrics get higher with the increase of the edge density. For instance, the uncertainly level on the selected team skill decreases and tends to zero when the edge density reaches one. This can be explained by the fact that, by increasing the edge density of the graph, the number of hops between the leader and any worker in $\mathcal G(\mathcal J, \mathcal E)$  decreases. Moreover, the number of directly connected workers increases until the resultant graph is fully connected (i.e., everyone knows everyone). Also, the team skills level, budget allocation, and relationships rate increase when increasing the edge density because the team leader will have more workers connected to it and consequently more vast choices in its vicinity. When increasing the edge density, the relationships rate within the team increases because team members are more likely to have more connections. However, the performances of the platform-based approach, except the relationships rate, remains basically invariant while varying the edge density. We notice a growth in the relationships rate and a slight decrease in all the other metrics. This can be explained by the fact that the platform is basing its recruitment decision on the workers' history and any changes of the edges density in $\mathcal G(\mathcal J, \mathcal E)$ will only affect the relationships term in the objective function. Notice that, for the leader-based approach, the recruited team skills level exceeds the one of the platform-based approach when the SN network becomes nearly fully-connected. These observation remains valid for $J=28$. In fact, the effect of expanding the network while maintaining the number of required team members enhanced slightly the performance of both recruitment strategies by increasing the team skill level and decreasing the recruiter uncertainty. We notice that, for the leader-based approach, the team skill level curve has a higher slop and converge faster than when  $J=14$.

In Fig.~\ref{2figures}, we present an example of the recruited teams using both recruitment strategies for $J=14$, $|\mathcal S_p|=4$. The figure shows that the leader-based approach recruits a congregated team (workers are close to each other in the SN) while the platform-based approach recruits a team relatively scattered but with higher skills.

\section{Conclusion}\label{sec5}
In this paper, we presented two recruitment strategies that form a team of skilled and socially connected workers in CMCS IoT systems. The platform-based approach exploits the platform knowledge to hire the team. The leader-based approach uses workers' knowledge about their SN neighbors to designate a leader that recruits the rest of the team. Results show a performance trade-off between the two strategies when varying the workers SN edge degree. The platform-based strategy recruits a more skilled team than the leader-based approach but with lower SN relationships and higher rewards. In our future work, we will focus on designing low complexity recruitment algorithms enabling real-time CMCS operations.

\bibliographystyle{IEEEtran}
\bibliography{references}

\begin{thebibliography}{10}
\providecommand{\url}[1]{#1}
\csname url@samestyle\endcsname
\providecommand{\newblock}{\relax}
\providecommand{\bibinfo}[2]{#2}
\providecommand{\BIBentrySTDinterwordspacing}{\spaceskip=0pt\relax}
\providecommand{\BIBentryALTinterwordstretchfactor}{4}
\providecommand{\BIBentryALTinterwordspacing}{\spaceskip=\fontdimen2\font plus
\BIBentryALTinterwordstretchfactor\fontdimen3\font minus
  \fontdimen4\font\relax}
\providecommand{\BIBforeignlanguage}[2]{{%
\expandafter\ifx\csname l@#1\endcsname\relax
\typeout{** WARNING: IEEEtran.bst: No hyphenation pattern has been}%
\typeout{** loaded for the language `#1'. Using the pattern for}%
\typeout{** the default language instead.}%
\else
\language=\csname l@#1\endcsname
\fi
#2}}
\providecommand{\BIBdecl}{\relax}
\BIBdecl

\bibitem{7491206}
A.~{Bader}, H.~{Ghazzai}, A.~{Kadri}, and M.~{Alouini}, ``Front-end
  intelligence for large-scale application-oriented internet-of-things,''
  \emph{IEEE Access}, vol.~4, pp. 3257--3272, June 2016.

\bibitem{7538981}
J.~{Dofe}, J.~{Frey}, and Q.~{Yu}, ``Hardware security assurance in emerging
  iot applications,'' in \emph{2016 IEEE International Symposium on Circuits
  and Systems (ISCAS)}, May 2016, pp. 2050--2053.

\bibitem{article5}
D.~C.~Brabham, ``Crowdsourcing as a model for problem solving: An introduction
  and cases,'' \emph{Convergence: The International Journal of Research Into
  New Media Technologies}, Feb 2008.

\bibitem{8982179}
A.~{Hamrouni}, H.~{Ghazzai}, M.~{Frikha}, and Y.~{Massoud}, ``A spatial mobile
  crowdsourcing framework for event reporting,'' \emph{IEEE Transactions on
  Computational Social Systems}, vol.~7, no.~2, pp. 477--491, Apr. 2020.

\bibitem{7875084}
X.~{Gan}, X.~{Wang}, W.~{Niu}, G.~{Hang}, X.~{Tian}, X.~{Wang}, and J.~{Xu},
  ``Incentivize multi-class crowd labeling under budget constraint,''
  \emph{IEEE Journal on Selected Areas in Communications}, vol.~35, no.~4, pp.
  893--905, April 2017.

\bibitem{7446292}
P.~{Cheng}, X.~{Lian}, L.~{Chen}, J.~{Han}, and J.~{Zhao}, ``Task assignment on
  multi-skill oriented spatial crowdsourcing,'' \emph{IEEE Trans. Knowl. Data
  Eng.}, Aug 2016.

\bibitem{8488386}
J.~{Flores-Parra}, M.~{Castañón-Puga}, R.~D. {Evans}, R.~{Rosales-Cisneros},
  and C.~{Gaxiola-Pacheco}, ``Towards team formation using belbin role types
  and a social networks analysis approach,'' in \emph{2018 IEEE Technology and
  Engineering Management Conference (TEMSCON)}, June 2018, pp. 1--6.

\bibitem{2222}
A.~{Lakhani}, A.~{Gupta}, and K.~{Chandrasekaran}, ``Intellisearch: A search
  engine based on big data analytics integrated with crowdsourcing and
  category-based search,'' in \emph{2015 International Conference on Circuits,
  Power and Computing Technologies [ICCPCT-2015]}, March 2015, pp. 1--6.

\bibitem{Mass1912}
A.~Hamrouni, H.~Ghazzai, T.~Alelyani, and Y.~Massoud, ``A stochastic team
  formation approach for collaborative mobile crowdsourcing,'' in \emph{IEEE
  International Conference on Microelectronics (ICM'19)}, Cairo, Egypt, 2019.

\bibitem{7806269}
S.~{Murali}, V.~{Krishnapriya}, and A.~{Thomas}, ``Crowdsourcing for disaster
  relief: A multi-platform model,'' in \emph{2016 IEEE Distributed Computing,
  VLSI, Electrical Circuits and Robotics (DISCOVER)}, Aug 2016, pp. 264--268.

\bibitem{article}
A.~Singla, E.~Horvitz, P.~Kohli, and A.~Krause, ``Learning to hire teams,'' in
  \emph{AAAI Conference on Human Computation and Crowdsourcing (HCOMP 2015)},
  San Diego, California, USA, Nov. 2015.

\bibitem{7248382}
Q.~{Liu}, T.~{Luo}, R.~{Tang}, and S.~{Bressan}, ``An efficient and truthful
  pricing mechanism for team formation in crowdsourcing markets,'' in
  \emph{2015 IEEE International Conference on Communications (ICC)}, June 2015,
  pp. 567--572.

\bibitem{AAAI1612106}
Z.~Pan, H.~Yu, C.~Miao, and C.~Leung, ``Efficient collaborative
  crowdsourcing,'' in \emph{AAAI Conf. Arti. Intel. (AAAI'16)}, 2016.

\bibitem{116}
H.~Jiang and S.~Matsubara, ``Efficient task decomposition in crowdsourcing,''
  in \emph{PRIMA 2014: Principles and Practice of Multi-Agent Systems}, H.~K.
  Dam, J.~Pitt, Y.~Xu, G.~Governatori, and T.~Ito, Eds.\hskip 1em plus 0.5em
  minus 0.4em\relax Cham: Springer International Publishing, 2014.

\bibitem{article7}
A.~Anagnostopoulos, L.~Becchetti, C.~Castillo, A.~Gionis, and S.~Leonardi,
  ``Online team formation in social networks.''\hskip 1em plus 0.5em minus
  0.4em\relax International Conference on World Wide Web (WWW'12), Lyon,
  France, Apr. 2012.

\bibitem{article8}
M.~Kargar and A.~An, ``Discovering top-k teams of experts with/without a leader
  in social networks,'' in \emph{ACM Int. Conf. Inf. Knowl. Manag.}, ser. CIKM
  '11, New York, NY, USA, 2011.

\end{thebibliography}

\end{document}